\documentclass[twocolumn]{aastex62}
\graphicspath{{./}{figures/}}

\received{}
\revised{}
\accepted{}
\submitjournal{ApJ}

\shorttitle{T CrB X-ray observations}
\shortauthors{Luna et al.}

\newcommand{\suzaku}{{\it Suzaku}}
\newcommand{\xmm}{{\it XMM}-Newton}

\newcommand{\swift}{{\it Swift}}

\newcommand{\nustar}{\textit{NuSTAR}}

\newcommand{\ms}{$M_{\odot}$}

\newcommand{\fluxcgs}{ergs~s$^{-1}$~cm$^{-2}$~\rm}
\newcommand{\lumcgs}{ergs~s$^{-1}$}
\newcommand{\chisq}{$\chi^{2}_{\nu}$}

\begin{document}

\title{Dissecting a disk-instability outburst in a symbiotic star: 
\nustar\, and \swift~ observations of T Coronae Borealis during the rise to the ``super-active''  state}

\correspondingauthor{Gerardo J. M. Luna}
\email{gjmluna@iafe.uba.ar}

\author{G. J. M. Luna}
\affil{CONICET-Universidad de Buenos Aires, Instituto de Astronom\'ia y F\'isica del Espacio (IAFE), Av. Inte. G\"uiraldes 2620, C1428ZAA, Buenos Aires, Argentina}
\affiliation{Universidad de Buenos Aires, Facultad de Ciencias Exactas y Naturales, Buenos Aires, Argentina.}

\author{T. Nelson}
\affiliation{Department of Physics and Astronomy, University of Pittsburgh, Pittsburgh, PA 15260}

\author{K. Mukai}
\affiliation{CRESST and X-ray Astrophysics Laboratory, NASA Goddard Space Flight Center, Greenbelt, MD 20771, USA}
\affiliation{Department of Physics, University of Maryland, Baltimore County, 1000 Hilltop Circle, Baltimore, MD 21250, USA}

\author{J. L. Sokoloski}
\affiliation{Columbia Astrophysics Lab 550 W120th St., 1027 Pupin Hall, MC 5247 Columbia University, New York, New York 10027, USA }
\affiliation{Large Synoptic Survey Telescope Corporation, 933 North Cherry Ave,
Tucson, AZ 85721.}



\begin{abstract}

The current {\em super-active} state of the recurrent nova T~CrB has been observed with unprecedented detail. Previously published observations provide strong evidence that this state is due to an enhancement of the flow of material through the accretion disk, which increased the optical depth of its most internal region, the boundary layer. \nustar\ and \swift\ observed T~CrB in 2015 September, roughly halfway through the rise to optical maximum.
In our analysis of these data, we have found that: $i$) the UV emission, as observed with \swift/UVOT in 2015, was already as bright as it became in 2017, after the optical peak; $ii$) the soft X-ray emission (E $\lesssim$ 0.6 keV) observed in 2017 after the optical peak, on the other hand, had not yet developed during the rising phase in 2015; $iii$) the hard X-ray emitting plasma (E $\gtrsim$ 2 keV) had the same temperature and about half the flux of that observed during quiescence in 2006. 
This phenomenology is akin to that observed during dwarf novae in outburst, but with the changes in the spectral energy distribution happening on a far longer time scale. 




\end{abstract}



\section{Introduction} 
\label{sec:intro}

Depending on the rate at which material
flows through
the accretion disk, the optical depth of its most internal, non-Keplerian region
can
be large or small. This region, known as boundary layer, would thus be optically thin or thick to its own radiation. A change in the accretion rate would change this condition. In dwarf novae (where a white dwarf accretes from a dwarf companion that orbits the white dwarf with period of hours), 
a thermal instability of the accretion disk changes the rate at which matter flows through the disk. On the rise to the outburst (optical high state), this typically happens in about 1 day, while during the decay, the transition usually takes several days. 
In dwarf novae, the transition from optically thin to thick regimes during outbursts manifests itself as an increase in the optical/soft-X-ray/EUV flux and a decrease in the hard X-rays \citep{2003MNRAS.345...49W}. 

In white dwarf symbiotic binaries, such as T~Coronae Borealis (T~CrB), a white dwarf accretes from a red giant companion
that orbits the white dwarf with a period of hundreds of days to years, a $\sim$AU-size accretion disk might form around the white dwarf (WD). Symbiotic stars often experience brightening episodes with changes of 2-3 magnitudes in the optical light curves, known as symbiotic classical outburst \citep[see e.g.][]{1986syst.book.....K}. Interpretations for the origin of such outbursts include WD photospheric expansion at constant luminosity due to an accretion rate that is above the value for stable surface nuclear-burning \citep{1976Afz....12..521T}; a shell flash \citep{1983ApJ...273..280K}; a disk-instability as in dwarf-nova outburst \citep{1986A&A...163...56D}; or some combination of these phenomena \citep{2006ApJ...636.1002S}. Because many WDs in symbiotic binaries are intrinsically very luminous in the optical--EUV due to quasi-steady shell burning on their surfaces, changes in the accretion disk itself are often difficult to observe directly.

T~CrB, however, is also a recurrent nova \citep{2010ApJS..187..275S}, indicating that accreted material does not burn quasi-steadily on the surface of the WD.  Nova eruptions (in which accreted hydrogen is suddenly ignited on the WD surface) took place in T~CrB in 1866 and 1946; the presence of more than one nova outburst within a century likely indicates that T~CrB hosts a massive WD accreting at a high rate from its red giant companion, although not high enough to produce quase-stable nuclear burning on the WD surface.  This lack of quasi-steady burning furnishes an important opportunity to diagnose the behavior of a WD accretion disk that is orders of magnitude larger than the WD disks in CVs.


In \citet{2018A&A...619A..61L}, we determined that a small optical brightening event ($\Delta V \sim$ 1) that started in early 2014 \citep{2016NewA...47....7M} was due to an increase in the rate of accretion through the disk, which caused the boundary layer to become optically thick. The observed phenomenology, which consisted of a softening of the X-ray spectrum and the appearance of a black-body component, a fading in the hard X-ray flux and an increase in the UV brightness confirmed, for the first time in a symbiotic binary, that this event was due to an increase in the accretion 
flow through the disk, similar to a disk instability dwarf nova outburst. 


In this paper, we present contemporaneous X-ray observations of T~CrB obtained with the \nustar\ and the {\it Neil Gehrels Swift} observatories. Both observations occurred during the rising phase of the recent optical brightening (see Fig. \ref{fig1}), 
after which most of the boundary layer became optically thick.  We use these observations to characterize the properties of the outburst, the time scale for the transition as manifested at various wavebands, and the accretion rate at which these occurred.  We also revisit the 2006 \suzaku\ spectrum (hereafter \suzaku$_{2006}$) to assess any long-term changes in the accretion rate and absorbing structures of the system.  Throughout the paper we assume a distance to T~CrB of 800$^{+30}_{-30}$ pc, from the parallax measurement published in GAIA DR2 \citep{2018AJ....156...58B}.

\section{Observations and Data Reduction}
\label{data}

\subsection{\textit{NuSTAR}}


We observed T~CrB with the \nustar\ satellite on 2015 September 23 for a total of 79.8 ks (hereafter \nustar$_{2015}$), well into the rising phase of the latest optical brightening event.  
We used the {\it nuproducts} v0.3.0 software to extract both spectral and light curve products, resulting in a spectrum and light curve for each FPMA and FPMB units. Source events were extracted from a 70 arcsec circular region centered on $\alpha$=15h 59m 30.160s and $\delta$=+25$^{\circ}$ 55$^{\prime}$ 12.59$^{\prime\prime}$,
while background events were extracted from an equally sized circular region off-source.  
The resulting spectra were binned to have a minimum of 25 counts per bin.

\subsection{\textit{Swift}}

The \swift\ satellite observed T~CrB twice during the \nustar$_{2015}$ observation (ObsId 00081659001 and 00081659002, hereafter \swift$_{2015}^{1}$) and a week later (ObsID 00045776003 and 00045776004, hereafter \swift$_{2015}^{2}$).  The total duration of the combined exposures of \swift$_{2015}^{1}$ was 9853 s and of \swift$_{2015}^{2}$ is 2953 s.  Data products 
were created by first combining the event files from the two observations and 
then extracting source events 
from a circular region of radius 20 pixels centered on the SIMBAD coordinates of T~CrB, and background events from an off-source circular region with a radius of 80 pixels.  
The ancillary response files were created using the {\tt xrtmkarf} tool.  

During these same observations, we also obtained UVOT images in the U ($\lambda$3465\AA, FWHM=785 \AA) and UVW2 ($\lambda$1938 \AA, FWHM=657 \AA) filters.  T~CrB was bright enough to saturate the UVOT detector, resulting in severe coincidence losses that mean that source brightness cannot be estimated using the standard photometry tools in HEAsoft.  Instead, we utilize the readout-streak in the images to estimate the source brightness \citep{2013MNRAS.436.1684P}. 


\begin{figure*}
\begin{center}
\includegraphics[scale=0.7]{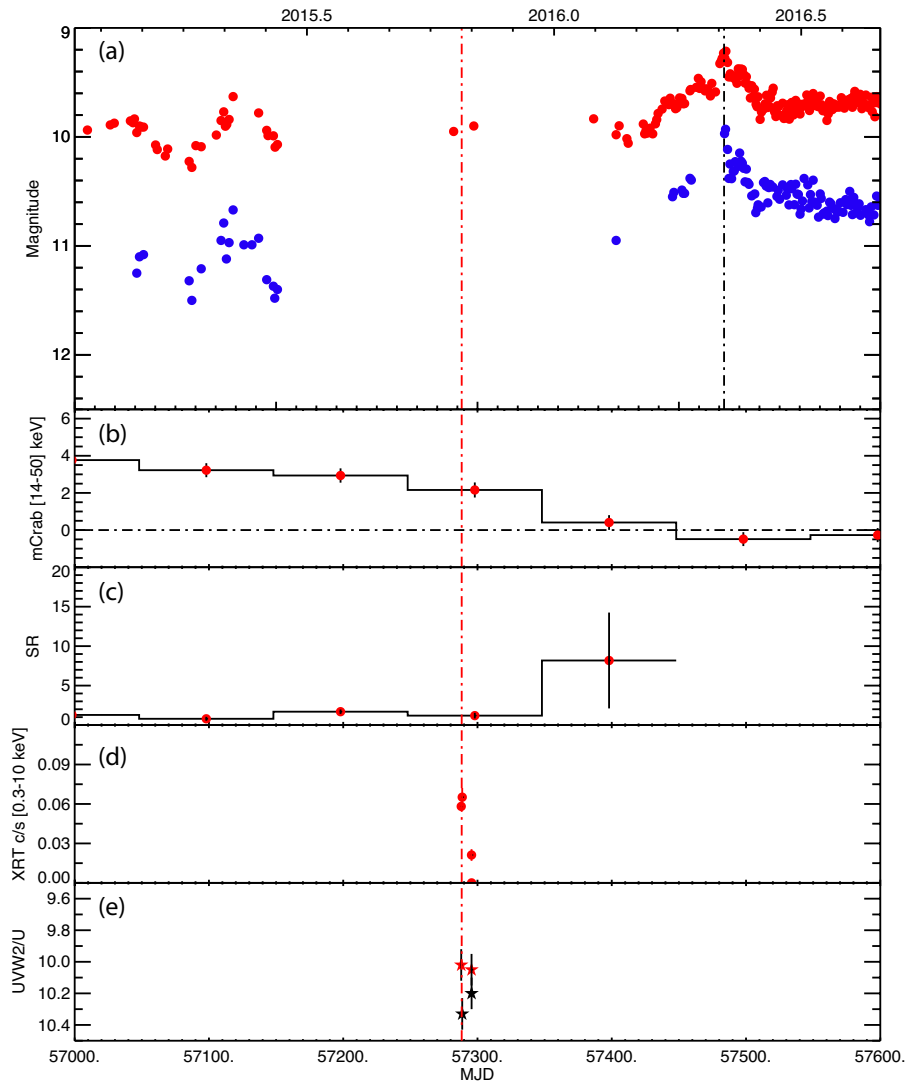}
\caption{{\it (a)}: T CrB AAVSO B (blue dots) and V (red dots) light curves.  Vertical, dashed lines show the date of the \nustar+\swift$^{1}$\ observation (red) and date of maximum optical brightness (black). {\it (b)} \swift\ BAT 14--50 keV light curve with 100 days bins. It is evident that the BAT flux started to decay quasi-simultaneously with the increase in the optical flux, which started around December 2014, it became too faint for its detection right around the optical maximum. {\it (c)} \swift\ BAT softness ratio (15--25/25--100 keV) with 100 days bins. This ratio steeply increased owing to the softening of the X-ray emission. {\it (d)} \swift\ XRT 0.3-10 keV count rate. (e) \swift\ UVOT UVW2 (red stars) and U (black stars) magnitudes determined from the CCD readout streak. Even during the rise to optical maximum, the UV flux has already increased dramatically, while previous to these measurements the UV (UVM2) magnitude was about 14. 
}
\label{fig1}
\end{center}
\end{figure*}

\section{Results.}
\subsection{ X-ray Spectral Analysis.}

\begin{figure*}
\begin{center}
\includegraphics[scale=0.5]{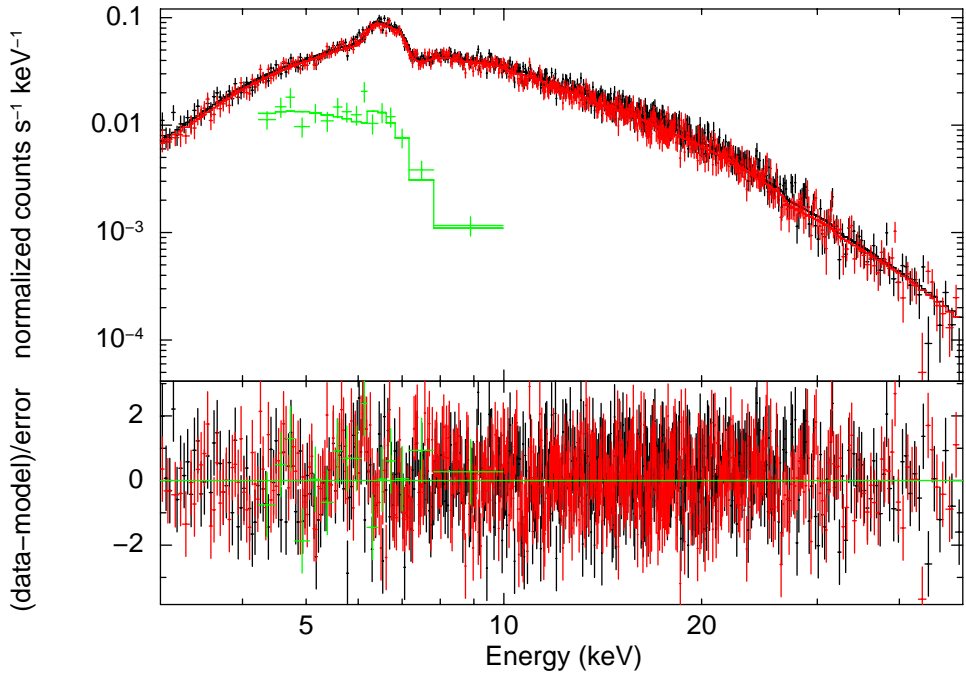}
\caption{\nustar$_{2015}$+\swift$_{2015}^{1}$ X-ray spectra of T CrB data with the best spectral model (solid line) which includes both a partial covering absorber and reflection, {\tt TBabs$\times$(partcov$\times$TBabs)$\times$(reflect$\times$mkcflow+gauss)} in XSpec.  }
\end{center}
\label{fig2}
\end{figure*}

The \nustar$_{2015}$ spectra are well suited to looking for evidence of reflection of X-rays from the white dwarf surface.  We explored possible models for T~CrB by jointly fitting the two \nustar$_{2015}$ modules and the \swift$_{2015}^{1}$/XRT data (Fig. \ref{fig2}). The short exposure time and low number of counts detected during the \swift$_{2015}^{2}$ observation did not provide a significant improvement to the joint fit. 
We fit the \nustar$_{2015}$ spectra in the range 3--78 keV, and the \swift$_{2015}^{1}$/XRT spectrum in the range 0.3--10 keV.
For models that include reflection we used the {\tt reflect} model in XSpec, which calculates the reflection of X-rays from neutral material using the method of \citet{1995MNRAS.273..837M}.  
The angle between the line of sight and the reflecting surface normal ($\theta$) is a key parameter in the reflection model, but \citet{1995MNRAS.273..837M} showed that cos($\theta$)=0.45 provided a reasonable approximation for a wide range of $\theta$ (18$^{\circ}$-76$^{\circ}$; see their Figure 5). For the specific case of the white dwarf surface reflecting X-rays from an equatorial boundary layer, $\theta$ will be close to 90 if the binary is very close to exactly face on (binary inclination $i\sim$0), while for an edge-on system, what we need is the average over azimuth changing from $\theta$=90$^{\circ}$ to 0$^{\circ}$ back to 90$^{\circ}$, with real systems being intermediate cases. \citet{1997MNRAS.288..649D} evaluated this average for SS Cyg (assumed $i\sim$37$^{\circ}$) and estimated the average $\theta$ to be close to 60$^{\circ}$ (so cos($\theta$)$\sim$0.5); for T CrB, with a higher binary inclination, this will be somewhat lower. However, given the relatively slow dependence of reflection on cos($\theta$) in this regime, we have fixed it to the Xspec default value of 0.45 in our analysis. To use the reflection model, we must extend the energy response over which XSpec evaluates the model. For T~CrB, we need to do this well beyond the \nustar$_{2015}$ response.  We chose 200 keV as the high energy cutoff, and model in 400 linear bins using the command {\tt energies extend ,200.0,400,lin} in XSpec. 

Given our knowledge of the current super-active state, which is most likely due to an enhancement of the accretion rate
through the disk and an increase of the optical depth of the boundary layer \citep{2018A&A...619A..61L}, we considered that both complex absorption and reflection could be at play.
Our preferred spectral model that fits the \nustar$_{2015}$+\swift$_{2015}^{1}$ spectrum 
includes both a partial covering absorber and reflection, i.e. {\tt TBabs$\times$(partcov$\times$TBabs)$\times$(reflect$\times$mkcflow+gauss)} in XSpec. 
We prefer this model over the others described below both on physical grounds and because of the better distribution of fit residuals revealed with the $runs$ test statistic (or Wald-Wolfowitz test; a non-parametric test used to check the randomness of the residuals, where a {\em run} is defined as a succession of one or more identical values of the data which are followed and preceded by a different value).  This model yields a shock maximum temperature of $kT_{max}$=40$\pm$3 keV and a reflection fraction of 1.1$\pm$0.2 (model ``Both" in Table \ref{tab:phot}). 

Using the more up-to-date absorption model {\tt TBabs} \citep{2000ApJ...542..914W}, we also fit the spectra using the model that  
\citet{2008ASPC..401..342L} fit to the \suzaku$_{2006}$ spectrum, 
i.e. {\tt TBabs$\times$(partcov$\times$TBabs)$\times$(mkcflow+gauss)}. In this model, X-rays from the accretion disk boundary layer are attenuated by two high column density absorbers, one of which partially covers the source. This complex absorbing system was required to explain the lack of X-rays below 2 keV, and to reproduce the high energy curvature observed in the XIS and HXD \suzaku$_{2006}$ spectra. 
We find a statistically acceptable fit to the \nustar$_{2015}$+\swift$_{2015}^{1}$ data with the \suzaku$_{2006}$ model; the best-fit parameters are shown in the first column (``Complex absorption only") of Table \ref{tab:phot}.  This model yields a very high column density for the partial covering absorber of 1.62 $\times$ 10$^{24}$ cm$^{-2}$. The high column densities on the order of $10^{24}$~cm$^{-2}$ and $10^{23}$~cm$^{-2}$ of the partial and full covering absorbers, respectively, imply that, for the photons created behind the absorber, only those above 10 keV are being transmitted.  This is the energy range precisely where reflected X-rays from the white dwarf surface are expected to be observed.  Furthermore, with a column density in excess of 10$^{24}$ cm$^{-2}$, the partial covering medium is itself optically thick enough to be a reflecting surface for X-rays.
The results of X-ray spectral fitting using models without an explicit reflection component thus suggest that significant reflection was indeed present in the spectrum. 

We therefore also tested simpler models in which the high energy X-ray spectral shape was due only to reflection of X-rays near the white dwarf surface, with no complex absorption.  We modeled this in Xspec as {\tt TBabs$\times$(reflect$\times$mkcflow+gauss)}.
In this scenario, the cooling flow component is reflected by material near the white dwarf surface, which modifies its shape above 10 keV.  Finally, the Gaussian component models the 6.4 keV Fe K$\alpha$ line, which is also due to reflection but is not included in the reflect model. The resulting fit is also statistically acceptable, with \chisq/dof = 1.06/1091.  The best fit maximum temperature of the cooling flow 
was $kT$ = 39$\pm$2 keV, and the accretion rate through the optically thin X-ray plasma was lower than the value returned from the model "Complex absorption only", $\dot{M}_{thin} = 4.8\pm 0.2 \times 10^{-10}\; M_{\odot}$~yr $^{-1}$.  The absorber had a column density of 3.2$\times$ 10$^{23}$ cm$^{-2}$, and the reflection fraction was 1.8$\pm$0.3.  We disfavor this model based on the results from the {\em runs} test when compared with the model that includes both reflection and absorption (see Table \ref{tab:phot}).


\renewcommand{\arraystretch}{1.3}
\begin{deluxetable*}{|l|ccc|cc|}
\tablecaption{ \label{tab:phot}
\nustar$_{2015}$+\swift$_{2015}^{1}$ and \suzaku$_{2006}$ spectral fits. Unabsorbed X-ray flux and luminosity, in units of 10$^{-10}$ \fluxcgs~ and 10$^{33}$ \lumcgs, respectively, are calculated in the 0.3-50 keV energy band. Elemental abundances are quoted in units of abundances from  \citep{2000ApJ...542..914W}. Luminosity and $\dot{M}$ are determined assuming a distance of 800 pc. Statistical errors are calculated at the 90\% confidence level. Systematic errors between \suzaku\ and \nustar\ are of about 5-10\% depending on the energy range \citep{2017AJ....153....2M}.}
\tablehead{
\colhead{}& \multicolumn{3}{|c|} {\nustar~+ \swift}& \multicolumn{2}{|c|} {\suzaku} }
\startdata
&Complex Absorption only & Reflection Only & Both & Complex Absorption only& Both\\
\hline
N$_{H}$ (ISM) [10$^{22}$ cm$^{-2}$] & 0.05 & 0.05 & 0.05 & 0.05 & 0.05 \\ 
N$_{H}$ (int.) [10$^{22}$ cm$^{-2}$] & 32$\pm$5 & 32$\pm$1 & \nodata & 27$\pm$2 & 22$\pm$3\\
N$_{H}$ (P.C.) [10$^{22}$ cm$^{-2}$] & 162$\pm$80 & \nodata & 43$\pm$3 & 49$\pm$ 5 & 32$\pm$3\\
Covering fraction & 0.4$\pm$0.1  & \nodata & 0.93$\pm$0.06 &  0.71$\pm$0.05 & 0.8$\pm$0.1\\
kT$_{\rm max}$ (keV) & 42$\pm$5 & 39$\pm$2  & 40$\pm$3 & 49$\pm$3 & 43$\pm$3\\
$\dot{M}_{thin}$ [10$^{-9}$ M$_{\odot}$ yr$^{-1}$] & 0.9$\pm$0.1  & 0.48$\pm$0.02 & 0.55$\pm$0.02 & 0.93$\pm$0.07& 0.77$\pm$0.05\\
EW$_{Fe K\alpha}$ (eV) &  & 191$\pm$2 & & & 138$\pm$2\\
Z/Z$_{\odot}$ & 1.00$\pm$0.01 & 1.01$\pm$0.01 & 1.04$\pm$0.09 & 1.04$\pm$0.08 & 1.03$\pm$0.09\\
Reflection fraction & \nodata & 1.8$\pm$0.3 & 1.1$\pm$0.2 &\nodata&  1.1 \\
 \chisq & 1.1 & 1.06 & 1.01 & 1.01 &1.06\\
 d.o.f & 1090 & 1091 & 1090 &1322&1321\\
 Flux  & 0.95$\pm$0.02 & 0.50$\pm$0.01 & 0.58$\pm$0.01 & 1.00$\pm$0.05 & 1.11$\pm$0.01 \\
 Luminosity  & 7.4 & 3.89 & 4.5 & 7.8& 8.7\\
 Runs statistic$^{a}$ & -2.50 & -1.22 & -1.16 & \nodata & \nodata \\
\enddata
\tablenotetext{a}{The Runs test evaluates if the fit residuals are randomly distributed above and below zero. We tested if the hypothesis that the residuals are randomly distributed can be rejected at the 5\% significance level, with higher values of abs(Runs) indicating the hypothesis can be rejected.}
\end{deluxetable*}



The \suzaku$_{2006}$ data alone cannot probe the presence of reflection due to the systematic uncertainties in the HXD background spectrum which are energy-dependent. 
We can however apply the knowledge about reflection we have derived from the \nustar$_{2015}$ data to the \suzaku$_{2006}$ spectrum. Fixing the reflection fraction to 1.1 (as we found for the \nustar$_{2015}$ data), we find a maximum temperature for the cooling flow in 2006 ($kT$=43$\pm$3 keV) that is very similar to that in 2015.  The normalization for this model was slightly higher than for the \nustar$_{2015}$ observation, suggesting a higher accretion rate through the optically thin portion of the boundary layer of $\dot{M}_{thin}$=7.7$\pm$0.5$\times$10$^{-10}$ \ms~ yr$^{-1}$. 


To summarize our spectral results: ($i$) the distribution of fit residuals, as measured by the {\it runs} test statistic, indicates that the model that includes reflection and complex absorption is preferable over other models;
($ii$) both the \nustar$_{2015}$ and \suzaku$_{2006}$ spectra can be described by highly absorbed cooling flow models. There is clear evidence of reflection in the \nustar$_{2015}$ spectrum, and including this source of emission in the \suzaku$_{2006}$ data resulted in acceptable model fits. 
The maximum temperature of the cooling flow was similar in 2006 and 2015, but the \suzaku$_{2006}$ data implies a slightly higher accretion rate through the optically thin part of the boundary layer in 2006.  


\subsection{UVOT Data Analysis}

First, we used the readout streak method 
on the \swift\ UVOT data to estimate count rates through the U and UVW2 filters (corrected for coincidence losses) of 0.220 and 0.070 c s$^{-1}$, or Vega system magnitudes of 10.1 and 10.33, respectively. We used the UVW2 values to estimate the broadband UV flux of T~CrB at the time of the \nustar$_{2015}$ observation, which we take as a proxy of its disk luminosity.

Previously, \citet{1992ApJ...393..289S} analyzed the {\sl IUE\/} observations of T~CrB from 1979 January 5 through 1990 February 9.  They estimated the 1250--3200\AA\ flux (hereafter F$_{UV}$) for each pair of SWP and LWP/LWR spectra, assuming a reddening of E${(B-V)}$=0.15, and showed that T~CrB was highly variable in the
UV during the {\sl IUE\/} era: F$_{UV}$ ranged from 6.9$\times 10^{-11}$
ergs\,cm$^{-2}$s$^{-1}$ (on 1989 August 1) to 1.33$\times 10^{-9}$
ergs\,cm$^{-2}$s$^{-1}$ (on 1983 May 1; see their Table 1). They measured an increase in F$_{UV}$ from 1.57$\times 10^{-10}$ to 3.02$\times 10^{-9}$
ergs\,cm$^{-2}$s$^{-1}$ between 1989 March 2 and April 8, so T~CrB is demonstrably
capable of a factor of almost two change in approximately one month.

These {\sl IUE\/} observations provided low-resolution, wide-band
spectroscopy covering the 1250--3200\AA\ range, and the \swift\ UVOT
data provide filter photometry in a small part of the {\sl IUE\/} band.
We downloaded archival {\sl IUE\/} data from MAST\footnote{https://archive.stsci.edu/iue/}, already reduced
and fluxed, and combined short and long-wavelength spectra that were
taken on the same day or on nearby dates. We folded them through
the UVOT filter effective area curve to predict the magnitude that
would have resulted, had \swift\ UVOT observations taken place during
the {\sl IUE\/} era. We also made our own estimates of F$_{UV}$ during the nineteen eighties, by de-reddening the {\sl IUE\/} spectra using E${(B-V)}$=0.15 and integrating
over the 1250--3200\AA\ range. Our values are similar to those found by
\citet{1992ApJ...393..289S}, although often $\sim$10\% lower. This may be
due to different reduction and/or calibration of the {\sl IUE\/} data,
or due to numerical differences in, e.g., the de-reddening process.
The only other caution regarding the results of \citet{1992ApJ...393..289S}
is that they used an assumed distance to T~CrB of 1300 pc, so their
luminosity values need to be reduced by a factor of approximately 2.6
for a direct comparison with ours.

\begin{figure*}
\begin{center}
\includegraphics[scale=0.8]{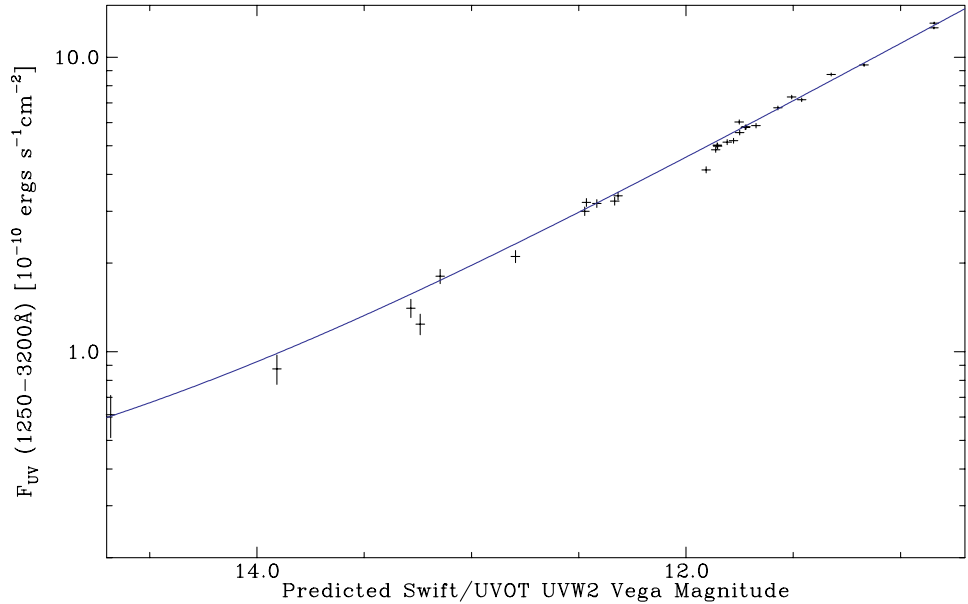}
\caption{ Predicted \swift\ UVOT magnitudes derived from IUE unabsorbed UV fluxes. }
\end{center}
\label{figfuvuvw2}
\end{figure*}

There are no major uncertainties in estimating equivalent \swift\ UVOT magnitudes from {\sl IUE\/} spectra, apart from the above mentioned
minor issues of exact reduction procedure and calibration. Going
in the other direction, from \swift\ UVOT photometry to F$_{UV}$
over the total {\sl IUE\/} range, on the other hand, involves
extrapolation of the source spectrum, which we do not directly
observe, and hence is highly uncertain.
It would be wrong to assume a typical accretion disk spectrum
(F$_\lambda \propto \lambda^{-2.33}$ power law) in the extrapolation:
\citet{1992ApJ...393..289S} showed that T~CrB had a highly variable
UV spectrum that was usually much flatter than the theoretical accretion
disk, except on one occasion when it had a power law index of 2.2.
However, we can convert \swift\ UVOT magnitudes to F$_{UV}$ by assuming the observed
{\sl IUE\/} spectra, collectively, are representative of the UV
spectra of T~CrB at the respective UV flux levels. Indeed, the
equivalent \swift\ UVOT (UVW2) magnitudes for {\sl IUE\/} data
show a tight correlation with F$_{UV}$ (see Figure \ref{figfuvuvw2}).
During the \swift$_{2015}$ observations, the UVW2 magnitudes were estimated
to be 10.33-10.20, which corresponds to UV (1250\AA~--3200\AA) fluxes of 20.4-23.0 (in units of 10$^{-10}$~ergs\,cm$^{-2}$~s$^{-1}$) or luminosities of 1.5-1.7$\times$10$^{35}$\lumcgs. These values should be interpreted with caution because given that T~CrB is so bright in the UV during the current high state, we are extrapolating the IUE-based UVW2 versus F$_{UV}$ relationship.
We use this estimate as our best (though imperfect)
proxy for the luminosity of the Keplerian disk.

\section{Discussion}

From the light curves presented in Figure \ref{fig1} we can identify a few different stages during the brightening phase. As mentioned in \citet{2018A&A...616A..53L}, in the optical, the brightening started early in 2014 and reached its maximum in 2016 April. We identify this as the rising phase from a mostly optically thin to a mostly optically thick boundary layer, 
which thus took about two years. During the optical rising phase, UV observations were taken along with the \nustar$_{2015}$ observation in 2015 September and a week later. These observations show that the UV flux had already risen to the value it attained during the 2016 optical high state by 2015 September, before the completion of the optical rise.
We identify this as the first sign that a heating wave had reached the inner disk. The next change appeared in the BAT light and softness ratio curves (panels $b$ and $c$ in Fig. \ref{fig1}), where a significant softening of the high energy emission started about 60 days after the \nustar$_{2015}$ (2015) observation and about 650-700 days after the beginning of the optical outburst. 

\begin{figure*}[ht!]
\begin{center}
\includegraphics[scale=0.8]{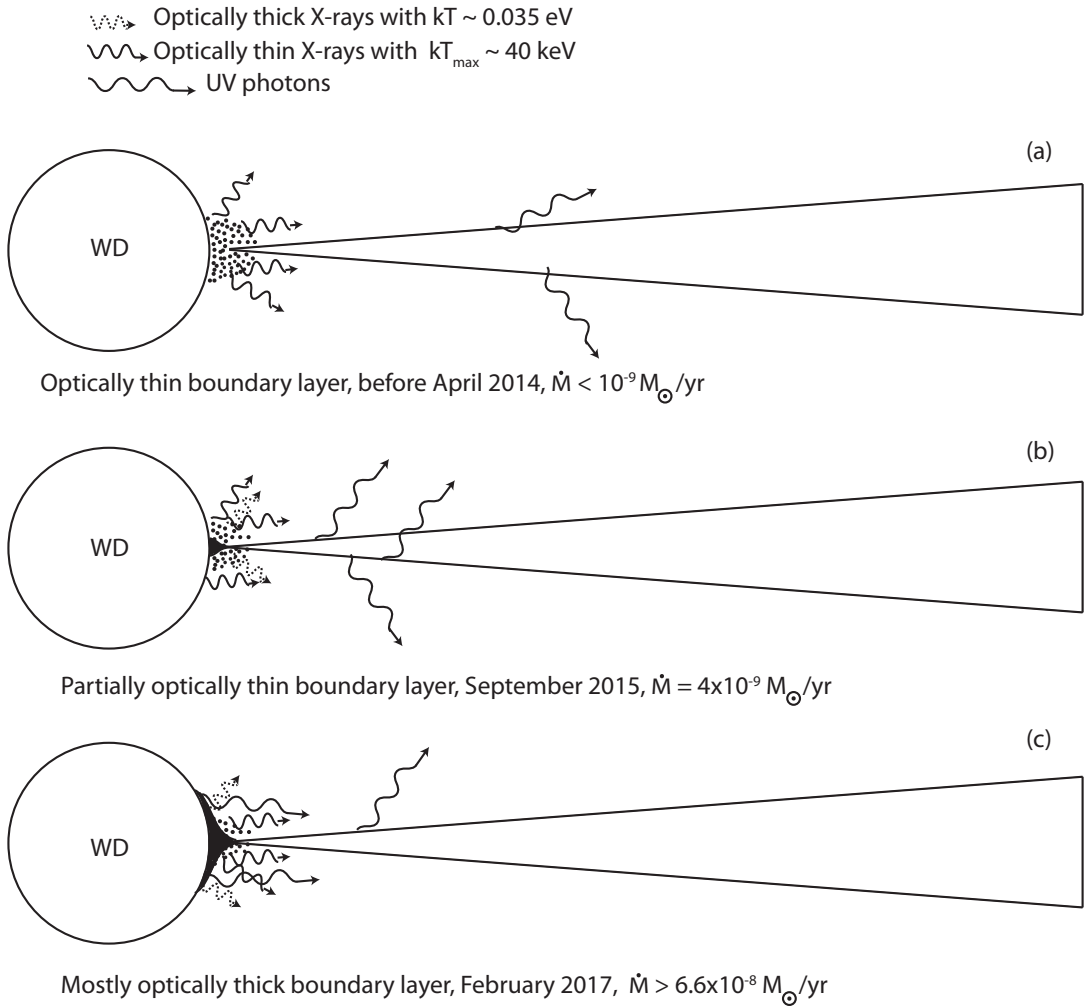}
\caption{Schematic view of the inner accretion disk and at the different stages that lead to the current {\em super-active} state. {\em (a)} Inner accretion disk and boundary layer state before April 2014, when $\log{\dot{M}} \lesssim$ 10$^{-9}$ \ms\ yr$^{-1}$. {\em (b)} State as of September 2015, with $\log{\dot{M}}$= 4$\times$10$^{-9}$ \ms\ yr$^{-1}$, when \nustar+\swift\ observations suggest that the boundary layer started its transition to the optically thick regime. {\em (c) Super-active} state, with most of the boundary layer being optically thick, as observed with \xmm\ in February 2017 \citep{2018A&A...616A..53L}.}
\end{center}
\label{fig:scheme}
\end{figure*}


The UV luminosity of T~CrB at the time of the \nustar$_{2015}$ observation was at least 5--10 times higher than the hard X-ray luminosity, depending on the X-ray spectral model we use (see Table \ref{tab:phot}). This ratio could have been even higher if there was a substantial bolometric correction (i.e., if a substantial fraction of disk luminosity was emitted shortward of 1250\AA).  The hard X-ray luminosity, on the other hand, represents that of the optically thin portion of the boundary layer. Thus, we find ourselves in the familiar position of trying to understand the missing boundary layer luminosity ``problem" \citep{1996A&A...315..467V}. 

Here we consider two broad categories of solutions for the missing boundary layer ``problem" in T~CrB. If the boundary layer was completely optically thin at the time of the NuSTAR observation in 2015, the accretion rate through the boundary layer must have been lower than that through the UV emitting portions of the accretion disk.  Alternatively, the boundary layer may already have been partially optically thick in 2015 September.
In the paragraphs that follow, we take these possibilities in turn.

If the entirety of the boundary layer of T~CrB was optically thin at the time of the \nustar$_{2015}$ observation, in which case its full luminosity would have been expected to appear in the hard X-rays, we need an explanation for the observational result that the boundary layer was significantly less luminous than the Keplerian disk at that time. Perhaps the brightening of T~CrB started in the outer disk (either due to an increased mass transfer rate from the donor or due to an outside-in disk instability), and the increase in the local accretion rate was propagating inward through the disk, similar to the familiar heating wave of the disk instability model, creating a delay between the UV and X-ray brightenings.  To quantify the magnitude of any such delay, we start by making a rough estimate of the location of the UV emitting region in the accretion disk. A 20,000 K blackbody has a spectral shape that peaks in the UV, and a spherical source with a radius of 
$10^{10}$~cm has a blackbody luminosity of just over 1.0$\times 10^{34}$ \lumcgs. The 
inferred UV luminosity of 3.5$\times 10^{34}$ \lumcgs\ therefore implies that the UV emission has a radius of less than 
$2 \times 10^{10}$~cm (a temperature significantly lower than 20,000 K would imply a larger radius for the same total luminosity, but much of the emission would shift into the optical range). In the case of a symbiotic system, this corresponds to the inner region of the disk, although comparable to the total size of a CV disk.
For this scenario to explain the low L$_{BL}$/L$_{disk}$ ratio during the optical rise, the heating wave must have reached the 
$R \sim 10^{10}$~cm region of the disk but not yet reached the boundary layer. In fact, judging by the later \swift$_{2015}^{2}$ observation as well as the low-time-resolution \swift\ BAT light curve, the delay between the heating wave reaching the inner part of the Keplerian disk and the boundary layer would have been at least 60 days. This is significantly longer than the duration of the transition we observe in dwarf novae, which takes about 2 days \citep[see e.g.][]{2003MNRAS.345...49W}.  Therefore we disfavor this scenario.

The alternative is that the boundary layer of T~CrB was partially optically thick at the time of the \nustar$_{2015}$ observation. In this scenario, we must explain why the X-ray characteristics of T~CrB were so different in 2015 (\nustar$_{2015}$) and 2017 (\xmm\ observations; hereafter \xmm$_{2017}$). The boundary layer of T~CrB was demonstrably fully optically thick in 2017, based on three characteristics \citep{2018A&A...619A..61L}. Most importantly, we directly observed a soft, optically thick, blackbody-like X-ray emission component. We also found that the hard X-ray flux was dramatically lower than in prior observations. Finally, the hard X-ray component was significantly softer than in prior observations, with $kT_{max}$ of about 13 keV.  The latter two characteristics of the boundary layer in 2017 are frequently seen in dwarf novae in outburst (see \citealt{2017PASP..129f2001M} for a review).  Given that the boundary layer was fully optically thick in 2017 and inconsistent with being fully optically thin in 2015, we favor the scenario in which it was partially optically thick in 2015.

With T~CrB showing evidence for a partially optically thick boundary layer in 2015, it is worth asking whether X-ray observations at that time revealed any direct evidence of soft, blackbody emission from the boundary layer or a decrease in the temperature of the hard X-ray component (both of which are associated with optically thick boundary layers).  Unfortunately, we were not sensitive to blackbody emission at the expected level. 
The upper limit on the luminosity of a blackbody component in 2015 (during the \nustar$_{2015}$ observation), based on \swift$_{2015}^{1}$ XRT data and taking the temperature we derived from the \xmm$_{2017}$ data, is 
$2 \times 10^{34}$ ergs\,s$^{-1}$. 
Whereas this upper limit on any blackbody X-ray emission in 2015 is more than an order of magnitude lower than the 2017 soft X-ray luminosity of L$_{bb}$=6.6$\times$10$^{35}$ \lumcgs (derived from the \xmm$_{2017}$ spectrum), it is of the same order as the inferred UV Keplerian disk luminosity in 2015.  And if the effective temperature in 2015 was lower than during the \xmm$_{2017}$ observation, an even higher luminosity soft component could have been hidden.   Regarding the hard X-ray component in 2015, however, we can clearly see that it was not significantly softer or fainter than that recorded in 2006 (\suzaku$_{2006}$).  T~CrB therefore showed very different behavior than dwarf novae in outburst.  In dwarf novae, the emergence of the soft X-ray component, UV brightening, 
the reduction in hard X-ray luminosity, and the softening of the hard X-ray spectrum all occur suddenly at a single transition point, as was famously seen during the rise to the outburst peak in the dwarf nova SS Cyg \citep{2003MNRAS.345...49W}.


Whereas the transition of the boundary layer from optically thin to thick during the rise to the outburst peak of dwarf novae appears sharp, the transition from thick to thin during the decay is slower and perhaps also more complicated \citep{2003MNRAS.345...49W}. We speculate that, perhaps, the optically thick boundary layer begins to emerge at one accretion rate, at which the characteristics of the residual hard X-rays are not immediately altered. The drop in hard X-ray luminosity and the softening of this optically thin component might happen at a higher accretion rate. Both changes appear to happen simultaneously during the rise of a dwarf nova outburst because the accretion rate through the boundary layer increases so sharply in a matter of hours.  In contrast, the accretion rate changes more slowly during outburst decay of dwarf novae, and in symbiotic stars, as demonstrated by the slow optical brightening of T~CrB.  In this interpretation, we caught the system at a very special time: after the initial emergence of UV emission from the optically thick boundary layer but before the optical depth had increased enough to have a dramatic effect on the properties of the hard X-ray component. 

\subsection{The transition accretion rate, $\dot{M}_{trans}$}

A major question for accreting white dwarfs, particularly those accreting at rates of around 10$^{-9}$ \ms\ yr$^{-1}$,
is the accretion rates at which the boundary layer transitions (or begins to transition) from optically thin to thick ($\dot{M}_{trans}^{1}$) regimes and vice versa ($\dot{M}_{trans}^{2}$). These transition accretion rates are a function of M$_{WD}$ and are closely tied to the assumed structure of the boundary layer.   Observational constraints on them can thus provide insight into boundary layer structure. Given that the time scale for the transition in one direction has been observed to be different from the time scale of the transition in the opposite direction in dwarf novae, it seems reasonable to suspect that the accretion rates for these transitions could also be different. 
However, only a few theoretical predictions of these thresholds exist.
For the purpose of comparing observations with theory, we return here to the simple picture in which all the empirical indicators of a boundary layer transition from optically thin to thick for a particular system appear at a single accretion rate $\dot{M}_{trans}^{1}$, and all the empirical indicators of a boundary layer transition from optically thick to thin appear at a distinct, single accretion rate $\dot{M}_{trans}^{2}$.
\citet{1993Natur.362..820N} concluded that when $\dot{M} = 3\times$10$^{-10}$ \ms\ yr$^{-1}$ and M$_{WD} = 1$~\ms, 
the boundary layer is generally expected to be optically thin, hot, and a source of hard X-rays, 
with a transition from this regime starting at $\dot{M}_{trans}^{1}\sim$10$^{-9}$ \ms\ yr$^{-1}$ (assuming a WD rotational speed of $\Omega$=0.5). At $\dot{M} \gtrsim \dot{M}_{trans}^{1}$, \citet{1993Natur.362..820N} found the boundary layer to be optically thick. 
%

Later, \citet{1995ApJ...442..337P} studied the optically thick solutions for the boundary layers in CVs and derived values of $\dot{M}_{trans}^{2}$ during the decay of an outburst
for different values of M$_{WD}$. By assuming that the transition would occur at $\tau_{*}$=1 (which includes opacity from free-free absorption from a fully ionized gas and electron scattering), the authors found that in a non-rotating WD, the transition 
would occur at a rate of 7.5$\times$10$^{-7}$ \ms\ yr$^{-1}$ (for a 1 \ms\ WD). Taking a definition of $\dot{M}_{trans}^2$ with a slightly smaller optical depth, $\tau_{*}$=0.8, yielded $\dot{M}_{trans}^{2}(\tau_{*}=0.8)$=4.6$\times$10$^{-8}$ \ms\ yr$^{-1}$ for the same WD mass. 
\citet{2014A&A...571A..55S} found that these thresholds might be overestimated due to an underestimation of the Rosseland opacity which 
might have previously been underestimated by two orders of magnitude.

Assuming that the WD mass can be estimated from the maximum cooling flow temperature in the X-ray spectral model that we refer to as ``Both", we derive a mass for the WD in T~CrB of at least
$1.15 \pm 0.03$ \ms;
in this case the theoretical models from \citet{1995ApJ...442..337P} predict an $\dot{M}_{trans}^{2}$ of about 4.6$\times$10$^{-8}$ \ms\ yr$^{-1}$ (taking $\tau$=0.8).
The observations with \xmm\ in 2017 (\xmm$_{2017}$) indicated an accretion rate of $\dot{M}\sim$ 6.6 $\times$10$^{-8}$ at that time  \citep{2018A&A...619A..61L}.  Observations during the end of the current state will therefore be crucial to test the theoretical prediction for $\dot{M}^{2}_{trans}$.

Observations of six dwarf novae compiled by \citet{2011PASP..123.1054F} during quiescence and outburst showed that the observationally-derived $\dot{M}^{1}_{trans}$ are much lower than the theoretical values predicted by \citet{1995ApJ...442..337P}. Moreover, the $\dot{M}^{1}_{trans}$ for these six dwarf novae are different from each other, ranging from $\sim$1.6$\times$10$^{-10}$ \ms\ yr$^{-1}$ in the case of SS~Cyg to $\sim$1.2$\times$10$^{-11}$ \ms\ yr$^{-1}$ in the case of U~Gem. Note, however, that \citet{2011PASP..123.1054F} implicitly assumed a single $\dot{M}^{1}_{trans}$, and took the highest recorded luminosity of the optically thin X-rays as corresponding to that $\dot{M}^{1}_{trans}$. If the emergence of an optically thick boundary layer and the decrease in luminosity of the optically thin X-ray component happens at different transition $\dot{M}$, this procedure does not provide an unambiguous result.  The behavior of U~Gem is unusual for a dwarf nova, and is a case in point: it has simultaneously exhibited a detectable soft component from the optically thick boundary layer and an enhanced optically thin X-ray component during
outburst \citep[see, e.g.][]{2000JAVSO..28..160M}.  During the 1993 December/1994 January outburst observed with EUVE three times, the soft component persisted to an estimated accretion rate of 3$\times$10$^{-10}$ \ms\ yr$^{-1}$ (scaled to a distance of 100 pc) so that could be taken as an indicator of $\dot{M}^{2}_{trans}$ \citep{1996ApJ...469..841L}.  On the other hand, the hard X-ray component does not disappear in U~Gem even when the accretion rate through the boundary layer exceed 5$\times$10$^{-9}$ \ms\ yr$^{-1}$, if all outbursts of U~Gem are similar to each other.

\subsection{The long-term accretion rate in T~CrB and implications for the next nova eruption.}

We observed an increase in accretion rate of several orders of magnitude between 2004 and 2018 \citep{2018A&A...619A..61L}.  Other evidence, primarily from IUE spectra obtained between 1979 and 1989, suggests that the accretion rate is generally variable by at least an order of magnitude \citep{1992ApJ...393..289S}.  Observations obtained with GALEX using the FUV ($\lambda$1528\AA; $\Delta\lambda$442\AA) filter in July 2006, show that the luminosity in the FUV filter was about L$_{FUV}$=2.2$\times$10$^{33}$ (d/800pc)$^{2}$  \citep[deriving the flux in the FUV band using \texttt{gPhoton};][]{2016ApJ...833..292M}. If this luminosity is roughly half of the available accretion luminosity, then the accretion rate at that time in 2006 was 2.2$\times$10$^{-10}$ \ms~ yr$^{-1}$ (and with L$_{BL}$/L$_{disk}\approx 1$, the boundary layer was optically thin). 
Theoretical nova outburst models such those presented by \citet{2005ApJ...623..398Y} predict that for a WD with a mass of about 1.0 \ms, accreting at $\sim$10$^{-8}$ \ms yr$^{-1}$, a nova outburst is produced every $\gtrsim$2$\times$10$^{3}$ years (see their Tables 2 and 3). The observed recurrence time of $\sim$80 years in T~CrB requires either a higher WD mass and average accretion rate, or a highly variable accretion rate. The episode of increased $\dot{M}$ reported in \citet{2018A&A...619A..61L} might not have been unique, and similar events might have passed unnoticed. 



\section{Conclusions}


The \nustar\ and \swift\ observations taken in 2015, during the rising phase of the current optical brightening allow us to estimate the accretion rate at which the boundary layer started to transition from an optically thin to an optically thick regime. This occurred at $\dot{M}_{trans}\approx 4 \times 10^{-9}$~(d/800 pc)$^{2}$ \ms~yr$^{-1}$.
We found that the soft X-rays observed after the optical maximum in \xmm$_{2017}$ had not yet appeared during the observations in 2015, discussed here. However, the UV emission had already become as bright as after the optical peak. The ratio of UV to X-ray luminosities, which we take to be a proxy for the optical depth of the boundary layer \citep{2013A&A...559A...6L}, was much greater than 1 in 2015.  Without the information provided by the later \xmm$_{2017}$ observation, we would have concluded that the boundary layer was mostly optically thick to its own radiation. However, the hard X-ray emission, which arises from the optically thin portion of the boundary layer, had not yet changed significantly.
Even though the accretion rate through the disk had increased enough by mid-2015 that the 
Keplerian portion shone mostly in UV, the boundary layer 
took at least several tens of days more to undergo all of the changes usually associated with the transition to high state in dwarf novae.
Our findings highlight how a key difference between symbiotic stars and cataclysmic variables -- the size of the accretion disk -- appears to impact phenomena close to the surface of the WD.


\acknowledgments

We acknowledge the anonymous referee for the useful comments that helped to improve the quality of this manuscript. We thank the {\it Swift} mission team for the generous allocation of target of opportunity time to observe T~CrB in support of the \nustar\ program.  We acknowledge with thanks the variable star observations from the AAVSO International Database contributed by observers worldwide and used in this research. This research has made use of the \nustar\ Data Analysis Software (NuSTARDAS) jointly developed by the ASI Science Data Center (ASDC, Italy) and the California Institute of Technology (Caltech, USA). This research has made use of the XRT Data Analysis Software (XRTDAS) developed under the responsibility of the ASI Science Data Center(ASDC), Italy. This work made use of data supplied by the UK \swift\ Science Data Centre at the University of Leicester. This work made use of the HEASARC archive. GJML is a member of the CIC-CONICET (Argentina) and acknowledge support from grants ANPCYT-PICT 0478/14, PICT 0901/2017, CONICET-NSF International Cooperation Grant 2016. JLS acknowledge support from the NSF through AST-1616646. 

\vspace{5mm}
\facilities{\nustar, \swift(XRT and UVOT), \suzaku}

\bibliography{listaref_MASTER}


\end{document}